# Above room-temperature two-dimensional ferromagnetic half-metals in Mn-based Janus magnets

Xiang-Fan Huang[1], Kang-Jie Li[2], Zequan Wang[1], Shi-Bo Zhao[1], Bing Shen[1], Zu-Xin Chen[2], Yusheng Hou [1,*]

**AFFILIATIONS**

[1] Guangdong Provincial Key Laboratory of Magnetoelectric Physics and Devices, Center for Neutron Science and Technology, School of Physics, Sun Yat-Sen University, Guangzhou, 510275, China

[2] School of Semiconductor Science and Technology, South China Normal University, Foshan, 528225, China

**ABSTRACT**

Two-dimensional (2D) ferromagnets and their heterostructures offer fertile grounds for designing fascinating functionalities in ultra-thin spintronic devices. Here, by first-principles calculations, we report the discovery of energetically and thermodynamically stable 2D ferromagnets with very strong inplane magnetic anisotropy in Mn$XY$ ($X$ = S, and Se; $Y$ = Cl, Br and I) monolayers. Remarkably, we find that the Curie temperatures of the ferromagnetic MnSBr, MnSI, MnSeCl, and MnSeI monolayers are as high as 271, 273, 231 and 418 K, respectively. In addition, we demonstrate that these ferromagnetic monolayers are intrinsic half-metals with large spin band gaps ranging from 2.5 eV to 3.2 eV. When spin-orbit coupling is considered in these ferromagnetic monolayers, the nature of their half-metal is almost unaffected. Finally, the strong inplane magnetic anisotropy of MnS$Y$ ($Y$ = Br, I) and MnSe$Y$ ($Y$ = Cl, I) monolayers originate mainly from halogen and chalcogen atoms, respectively. Our work shows 2D Janus Mn-based ferromagnetic half-metals may have appealing functionalities in high-performance spintronic applications.

Authors to whom correspondence should be addressed: houysh@mail.sysu.edu.cn

The experimental discoveries of atomically thin two-dimensional (2D) intrinsic magnets, such as CrI$_3$ monolayer (ML),[1] Cr$_2$Ge$_2$Te$_6$ bilayer,[2] and Fe$_3$GeTe$_2$ few layer,[3] have sparked tremendous scientific and technological interest, thanks to their unique potential applications in quantum computation,[4,5] logical and memory operation,[6] and spintronic devices.[7] Particularly, 2D ferromagnetic (FM) half-metals which are metallic in one spin channel but insulating in the other, are good sources of spin-flow injection and can meet the growing demands of high-performance spintronic devices due to their 100% spin polarization at the Fermi level.[8] Up to now, many 2D FM half-metals have been predicted in previous studies, such as Mn$X$ ($X$ = P, As),[9] Fe$X_2$ ($X$ = Cl, Br, I),[10] Fe$XY$ ($X, Y$ = Cl, Br and I; $X \neq Y$),[11] and Cr$_2$N$X_2$ ($X$ = O, F, OH).[12] Nevertheless, in most of the previously studied 2D FM half-metals, the unstable FM ground state at ambient temperatures, low Curie temperature ($T_C$) and weak magnetic anisotropy hinder their experimental verifications and practical applications. Hence, it is highly desirable to rationally design suitable 2D intrinsic FM half-metals with high $T_C$ and strong magnetic anisotropy.

Over the past few years, many endeavors have been made to realize robust 2D ferromagnets with high $T_C$ and strong magnetic anisotropy. One of the proven ways is modulating magnetism in existing 2D layered materials by post-treating, such as strain engineering,[13,14] defect engineering,[15,16] charge-doping,[17,18] and Janus engineering.[19,20] Among these strategies, Janus engineering is widely used and quite successful. This method is first proposed theoretically by Cheng et al.[21] Four years later, it is confirmed experimentally by Lu et al.[22] who successfully synthesized nonmagnetic Janus MoSSe ML. Immediately, this breakthrough inspired a surge of experimental and theoretical studies on 2D Janus magnets.[23] For example, it is shown that the $T_C$ of Janus VSSe ML can be as high as 400 K and its stability is comparable to the pristine VSe$_2$.[24] Utilizing Janus engineering, some stable FM semiconductors are predicted, such as Janus Cr$SX$ and Cr$_2X_3$ ($X$ = Cl, Br and I) MLs.[25,26] Besides, ferromagnetism and half-metallicity are demonstrated to coexist in Janus Fe$XY$ ($X, Y$ = Cl, Br and I, $X \neq Y$) MLs, but their $T_C$ are very low (below 30 K).[11] Although Janus FM half-metal Mn$_2X$Sb ($X$ = As, P) MLs have a $T_C$ higher than the room temperature, their magnetic anisotropy are very weak.[27]

Considering the high tunability of the composition elements in Janus magnets, we here explore the possibility of achieving 2D FM half-metals with a high $T_C$ and strong magnetic anisotropy.

In this work, utilizing first-principles calculations, we systematically investigate the electronic and magnetic properties of 2D Janus Mn$XY$ ($X$ = S and Se; $Y$ = Cl, Br and I) MLs. We demonstrate that MnSBr, MnSI, MnSeCl and MnSeI MLs have Jahn-Teller (JT) distortions and are thermodynamically stable but MnSCl and MnSeBr are unstable. Excitingly, the four stable MLs are desired intrinsic ferromagnets with strong inplane magnetic anisotropy and the $T_C$ of MnSeI is 418 K, much higher the room temperature. Additionally, these four MLs are intrinsic FM half-metals with large spin band gaps over 2.4 eV. When spin-orbit coupling (SOC) is considered, the 100% spin polarization at the Fermi level is not changed for MnSBr whereas the spin polarization at the Fermi level of other three MLs will be lower than 100%. Finally, we unveil that the strong inplane magnetic anisotropy of MnS$Y$ ($Y$ = Br, I) MLs originates mainly from halogen atoms while that of MnSe$Y$ ($Y$ = Cl, I) MLs comes mainly from chalcogen atoms. Our work introduces Mn-based magnets into the 2D FM half-metal family and provides promising platforms for designing high-temperature spintronic devices.

Our first-principles calculations based on density functional theory (DFT) are carried out using the Vienna *ab initio* simulation package (VASP) with the generalized gradient approximation.[28,29] We employ projector-augment wave pseudopotentials to describe core-valence interactions.[30,31] To describe the strong correlations between 3$d$ electrons of Mn, we adopt $U$ = 5.0 eV and $J$ = 1.0 eV. To avoid the spurious interactions between adjacent Mn$XY$ MLs along the normal axis, a vacuum space of 15 Å is added. Phonon spectra are calculated with the finite displacement method.[32,33] The *ab initio* molecular dynamics (AIMD) simulations use the NVT canonical ensemble and Nosé–Hoover thermostat.[34,35] Based on the DFT calculated magnetic parameters, we perform Monte Carlo (MC) simulations with the Metropolis algorithm to explore the magnetic ground states of Mn$XY$ MLs.

Fig. 1(a) shows the crystal structure of Mn$XY$ ML. One can see that Mn atom layer is sandwiched by chalcogen atom $X$ ($X$ = S and Se) and halogen atom $Y$ ($Y$ = Cl, Br and

I) layers. For individual Mn atom, it sits at the center of a distorted octahedron which is formed by ligand atoms *X* and *Y* (Fig. 1(b)). Because of the octahedral crystal field generated by $X^{2-}$ and $Y^{-}$ anions, the five-fold degenerate 3*d* orbitals of $Mn^{3+}$ cation are split into low-lying threefold degenerate $t_{2g}$ orbitals and high-lying twofold degenerate $e_g$ orbitals. Since $Mn^{3+}$ cations in Mn*XY* ML have localized 3*d* shells with a high-spin $d^4$ configuration, there are three electrons in $t_{2g}$ state and one in $e_g$ state, indicating that $Mn^{3+}$ cations are JT active.[36] By considering the JT distortion explicitly, we obtain that Mn*XY* MLs have a space group *Cm*, instead of the space group *P3m1* which is reduced from the space group $D_{3d}$ of Mn$Y_2$ (*Y* = Cl, Br and I) MLs.[37] Actually, as shown in Fig. S1 in supplementary material, we find that the phonon spectra of all Mn*XY* MLs with the space group *P3m1* have imaginary frequencies, which indicates that they are indeed unstable dynamically.

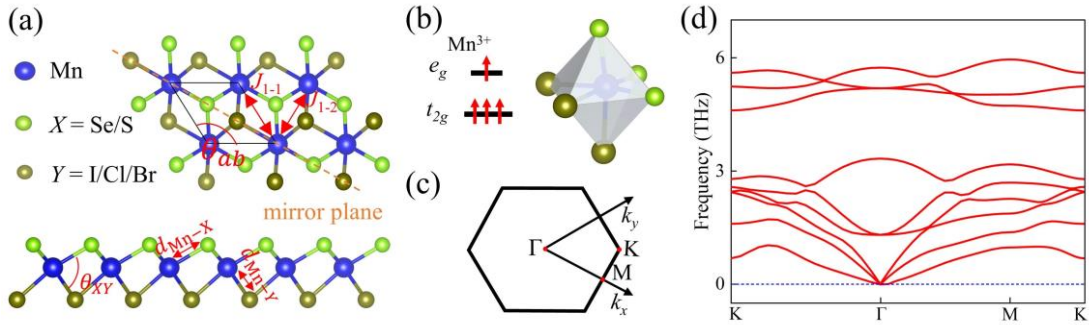

**FIG. 1.** (a) Top and side views of Mn*XY* ML. The orange dashed line presents the plane of mirror symmetry in Mn*XY* ML. (b) The left panel shows $3d^4$ configuration of $Mn^{3+}$ ion, and the right panel is the distorted Mn$X_3Y_3$ octahedron. (c) The first Brillouin zone of Mn*XY* ML and the high-symmetry points. (d) The phonon spectrum of MnSeI ML.

Here, we study the stability of Mn*XY* ML with the space group *Cm*. As an example, Fig. 1(d) shows the phonon spectrum of MnSeI ML. We see that there are indeed no imaginary frequencies. So, MnSeI ML with the space group *Cm* is dynamically stable. The phonon spectra of other five MLs are shown in Fig. S2 in supplementary material. Similar to MnSeI ML, MnSBr, MnSI and MnSeCl MLs have no imaginary frequencies in their phonon spectra, suggesting that they are also dynamically stable. For MnSeBr

ML, its phonon spectrum shows small negative frequencies in its out-of-plane acoustic phonon branch around the Γ point (Fig. S2(a) in supplementary material). This feature is usually associated with the structural instability caused by the in-plane bending of two different atom layers (i.e., Se and Br atom layers in MnSeBr), which is also reported in MnSSe[38] and some nonmagnetic Janus MLs.[39] For MnSCl ML, its phonon spectrum has large imaginary frequencies at the high-symmetry point M, which means that it is intrinsically unstable with the space group *Cm*. Besides, our AIMD simulations (Fig. S3 in supplementary material) suggest that MnSeI, MnSBr, MnSI and MnSeCl MLs are thermally stable. Thus, MnSBr, MnSI, MnSeCl and MnSeI MLs with the space group *Cm* is thermodynamically stable. Henceforth, we focus on these four stable MLs.

Now, let us have a deeper insight into the structural properties of the stable Mn*XY* MLs with the space group *Cm*. First of all, the symmetry of the space group *Cm* is very low and it has only three symmetry operations, namely, one identity operation, one glide symmetry operation and one mirror symmetry operation. As an important result of the low symmetry, the six nearest-neighbor (NN) Heisenberg exchange interactions end up being of two kinds, defined as $J_{1-1}$ and $J_{1-2}$, as shown in Fig. 1(a). A similar pattern is also observed for other Heisenberg exchange interactions and Dzyaloshinskii–Moriya interactions (DMIs) (Fig. S4 in supplementary material). As ligand atoms *X* and *Y* vary, Mn*XY* MLs have different in-plane lattice constants, bond lengths, bond angles and JT distortions (see Table S1 in supplementary material). Particularly, the JT distortion has a significant influence on the inplane lattice: the stronger the JT distortion is, the more the angle, $\theta_{ab}$, between the inplane *a* and *b* axes deviates from 120º. As a result of such deviation, the first Brillouin zone of Mn*XY* MLs is a distorted hexagon [Fig. 1(c)].

Having established the crystal structures of Mn*XY* MLs, we proceed to study their magnetic properties. The calculated net magnetic moment is 4 $\mu_B$/f.u. for Mn*XY* MLs, which is consistent with the high-spin $3d^4$ configuration of $Mn^{3+}$ cations. From Table S3 in supplementary material, we see that the local magnetic moments of $Mn^{3+}$ cation in Mn*XY* MLs are all over 4 $\mu_B$/Mn, predominantly contributing to the total magnetic moments. Due to their hybridizations with $Mn^{3+}$ cations, ligand anions $X^{2-}$ and $Y^-$ have nonzero induced magnetic moments which are antiparallel to the magnetic moments of $Mn^{3+}$ cations. Contrast to the small magnetic moments of 0.03-0.14 $\mu_B$/atom for ligand anions $Y^-$, ligand anions $X^{2-}$ have a bigger induced magnetic moment of 0.24-0.42 $\mu_B$/atom. Notably, the induced magnetic moments of $I^-$ anions in MnSI ML are as large as 0.14 $\mu_B$/atom, which is even larger than those of $I^{2-}$ anions in $CrI_3$.[40,41] The large magnetic moments of the ligand anions in Mn$XY$ MLs suggest they may have a strong magnetic proximity effect when contacting with other 2D functional materials.

To accurately determine the magnetic ground state of Mn$XY$ MLs, we adopt a spin Hamiltonian consisting of Heisenberg exchange interactions, DMIs and the single ion anisotropy (SIA). This spin Hamiltonian is in the following form:

$$H = \sum_{ij} J_{ij} \mathbf{S}_i \cdot \mathbf{S}_j + \sum_{<ij>} \mathbf{D}_{ij} \cdot (\mathbf{S}_i \times \mathbf{S}_j) + A \sum_i (S_i^z)^2 \quad (1).$$

In Eq. (1), $J_{ij}$ is the Heisenberg exchange parameter between spins $\mathbf{S}_i$ and $\mathbf{S}_j$, $\mathbf{D}_{ij} = (D_x, D_y, D_z)$ is the NN DMI vector and $A$ represents the SIA parameter. Here, negative and positive $J_{ij}$ mean FM and antiferromagnetic (AFM) Heisenberg exchange interactions, respectively. The Heisenberg exchange parameters $J_{ij}$ are calculated by the least squares fit technique[42] and four-state method is used to obtain the DMI vector.[43] Typically, DMI is short-range and usually dominated by the interactions between NN magnetic pairs.[44] Thus, only the NN DMI is considered in the following analysis. For the convenience of discussion, the magnitude and in-plane component of the NN DMI vector are defined as $|\mathbf{D}| = \sqrt{D_x^2 + D_y^2 + D_z^2}$ and $D_{//} = \sqrt{D_x^2 + D_y^2}$, respectively. Lastly, positive (negative) SIA parameter $A$ indicates an in-plane (out-of-plane) magnetic easy axis. The magnetic anisotropy energy (MAE) originating from the SIA is calculated by $E_{MAE} = E_{001} - E_{100}$, where $E_{001}$ and $E_{100}$ are the energies when the magnetic moments of Mn$XY$ MLs are long $z$ and $x$ axes, respectively.

As shown in Fig. 2(a), the Heisenberg exchange interactions $J_{i-1}$ and $J_{i-2}$ ($i$ = 1, 2 and 3) are almost FM in the four Mn$XY$ MLs. We also calculate $J_{4-j}$ and $J_{5-j}$ ($j$ = 1 and 2) and list them in Table S4 in supplementary material. For MnS$Y$ MLs, the values of $J_{i-1}$ are much larger than those of $J_{i-2}$ ($i$ = 1 and 2), while there is no such large difference between the values of $J_{i-1}$ and $J_{i-2}$ for MnSe$Y$. This discrepancy may originate from the distinct spacing between the two adjacent $Mn^{3+}$ cations corresponding to $J_i$ ($i$ = 1 and 2). For example, the two kinds of the NN distances for MnSBr ML are 3.54 Å and 3.75 Å, while these are 3.607 and 3.608 Å for MnSeCl ML. As illustrated in Fig. 2(b), all Mn$XY$ MLs have positive SIAs, suggesting that they all have an in-plane magnetic easy axis. Explicitly, the MAEs of MnSBr, MnSI, MnSeCl, and MnSeI MLs are 0.11, 0.42, 0.95 and 1.09 meV/Mn (Table S6 in supplementary material), respectively. In particular, MnSe$Y$ ($Y$ = Cl, I) MLs have much stronger MAEs than the FM half-metal $Mn_2AsSb$ ML (0.45 meV/Mn),[27] suggesting that they may have applications in magnetoelectronic devices.[45]

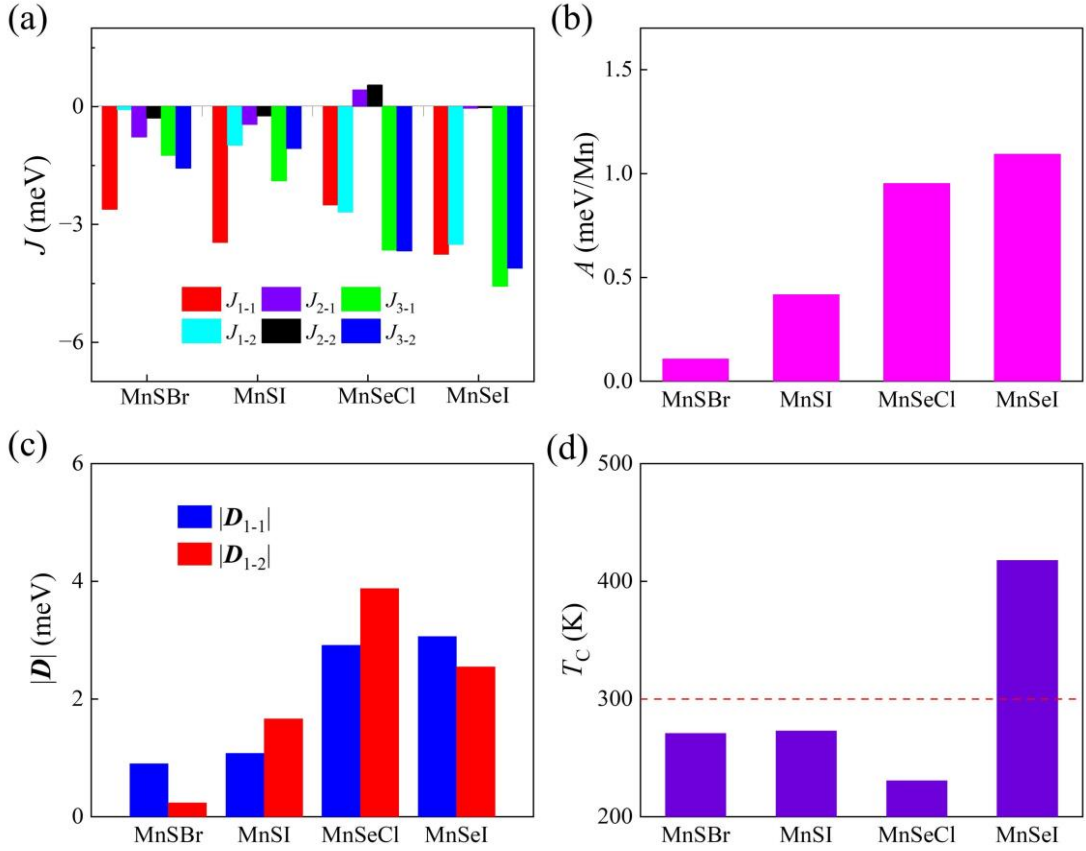

**FIG. 2.** Magnetic parameters of the four stable Mn$XY$ MLs. (a) Heisenberg exchange parameters $J_{i\text{-}j}$ ($i$ = 1, 2, 3; $j$ = 1, 2). (b) SIA parameter $A$. (c) $|D_{1\text{-}1}|$ (blue bar) and $|D_{1\text{-}2}|$ (red bar) of the NN DM interaction vector $D_1$. (d) MC simulated $T_C$.

Due to the inherent inversion symmetry breaking, Janus magnets often have strong

DMIs. As shown in Fig. 2(c), there are indeed significant DMIs in the four stable Mn$XY$ MLs. Since DMI is closely correlated with the SOC of ligand anions which are located between two adjacent magnetic cations in general, the magnitude of the DMI vector increases as the ligand anions $X^{2-}$ ($Y^-$) changes from $S^{2-}$ ($Cl^-$) to $Se^{2-}$ ($I^-$). It is noteworthy that the $|\boldsymbol{D}_{1-j,//}|/|J_{1-j}|$ ($j$ = 1 and 2) of MnSeI ML (Table S6 in supplementary material) is in the typical range of 0.1-0.2 which is known to generate magnetic skyrmions.[46] This suggests that MnSeI ML may be a candidate for realizing magnetic skyrmions.

Excitingly, our MC simulations show that the four stable Mn$XY$ MLs are high-temperature ferromagnets. As shown in Fig. 2(d), the $T_C$ of MnSBr, MnSI, MnSeCl and MnSeI MLs are as high as 271, 273, 231 and 418 K, respectively. Particularly, the $T_C$ of MnSeI ML is much higher than the room temperature. We notice that the values of the dominant $J_1$, $J_3$ and MAE in MnSeI ML are the largest among the four Mn$XY$ MLs, which thus lead to the highest $T_C$. Additionally, although MnSeI ML has sizable DMIs, its FM Heisenberg exchange interactions and in-plane MAE are strong enough to ensure the formation of FM ground states, according to our recently proposed descriptor linking the presence of magnetic skyrmions to magnetic parameters.[47] Compared with the low $T_C$ of the widely studied 2D ferromagnets, such as CrI$_3$ (45 K)[1] and Fe$_3$GeTe$_2$ (130 K)[3] MLs, the high $T_C$ of Mn$XY$ MLs are essential to their potential applications in practical spintronic devices.

Next, we investigate the electronic properties of Mn$XY$ MLs. Here, we take MnSeI ML as an example owing to its highest $T_C$ and show its spin-polarized band structure in Fig. 3(a). One can see that its spin-up channel crosses the Fermi level, whereas its spin-down channel is insulating with a band gap of 2.58 eV. As illustrated in Fig. 3(b), the projected density of state (PDOS) of MnSeI ML reveals that its electronic states near the Fermi level are contributed by Mn, Se and I atoms in its spin-up channel. As for its spin-down channel, the conduction band minimum is contributed mainly by Mn atoms, while the valence band maximum comes mainly from Mn and Se atoms. According to the definition of spin polarization,[48] it is calculated as $P = \left| \dfrac{N_\uparrow(E_F) - N_\downarrow(E_F)}{N_\uparrow(E_F) + N_\downarrow(E_F)} \right|$, where $N_↓(E_F)$ and $N_↑(E_F)$ are the spin-down and spin-up density of states (DOSs) at the Fermi level, respectively. Since $N_↓(E_F) = 0$, MnSeI ML is a FM half-metal with 100% spin polarization. As shown in Fig. S5 in the supplementary material, MnSBr, MnSI and MnSeCl are FM half-metal with 100% spin polarization as well, and the band gaps in their spin-down channels are 3.19, 2.69 and 2.49 eV, respectively. The 100% spin polarization and large band gaps in the spin-down channels of Mn$XY$ MLs indicate they may have wide applications for spintronic devices such as spin filters, spin injectors and magnetic sensors.

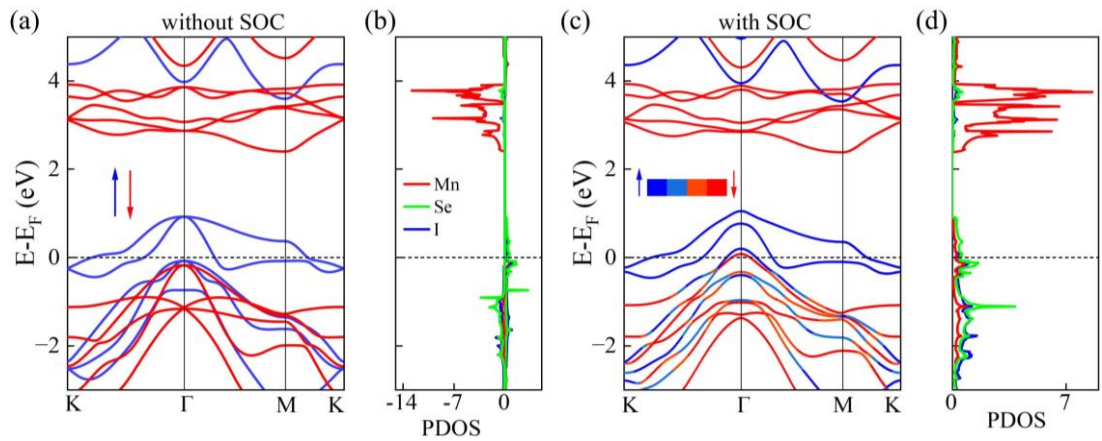

**FIG. 3.** Band structures and PDOS of MnSeI ML (a) without SOC and (b) with SOC. Up-spin and down-spin states are shown by blue and red colors in the band structures.

Considering the strong SOC of chalcogen atoms such as Se and halogen atoms such as I, we examine the effect of their SOC on the spin polarization of Mn$XY$ MLs. Due to their strong in-plane MAEs, we have the magnetic moments of $Mn^{3+}$ cations along the $x$ axis when calculating their band structures with SOC being considered. By comparing their band structures with and without SOC (Fig. 3(c) and Fig. S5 in supplementary material), we observe that SOC have no obvious effect on the overall profiles of band structures of the four Mn$XY$ MLs. However, SOC can induce noticeable changes for their bands at the Γ point near the Fermi level. For MnSI, MnSeCl and MnSeI MLs, their valence band maxima in the spin-down channels are pushed over the Fermi level at the Γ point. As a result of such changes, their $N_\downarrow(E_F)$ is no longer zero and thus, their spin polarizations are lower than 100%. It is worth noting that the SOC induced decreasing of spin polarization is also shown in zincblende half metals by a previous first-principles study[49] and non-100% spin polarization is experimentally determined in a representative half-metal, $CrO_2$.[50] As for MnSBr ML, its valence band maximum in its spin-down channel is not pushed over the Fermi level, so its 100% spin polarization is maintained. Overall speaking, the nature of the spin polarization near the Fermi level in Mn*XY* MLs is basically not changed by their SOC.

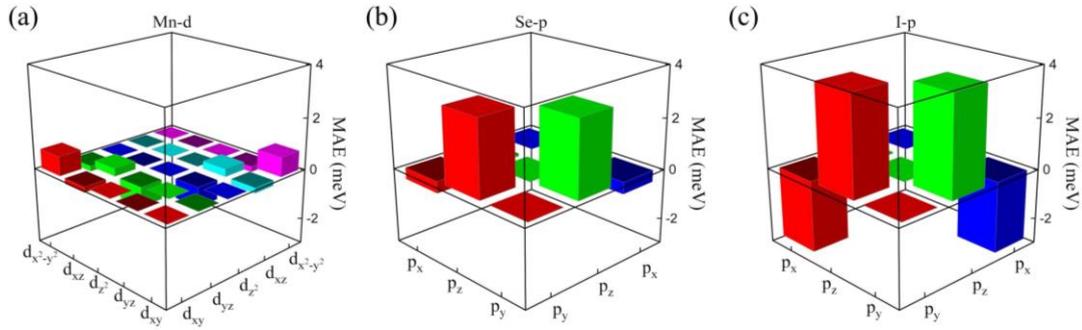

**FIG. 4.** The atomically orbital resolved MAE contribution of MnSeI ML from (a) *d* orbital hybridizations of Mn, (b) *p* orbital hybridizations of Se and (c) *p* orbital hybridizations of iodine.

Finally, we perform atomically orbital-resolved MAE to elucidate the underlying mechanisms of the strong inplane MAE in Mn*XY* MLs. As shown in Fig. 4 and Fig. S6 in supplementary material, we otain that the orbital hybridizations of Mn atoms make very small contributions to the MAEs in MnSBr, MnSI, MnSeCl and MnSeI MLs. By contrast, their ligand toms make the dominant contrition to their MAEs. As illustrated in Fig. 4(c) for MnSeI ML, the MAE contributions from the hybridizations between $p_z$ and $p_y$ orbitals of iodine atoms are positive while those from the hybridizations between $p_x$ and $p_y$ orbitals are negative. Consequently, the total MAE contribution from iodine atoms is also small. As the positive MAE contributions from the hybridizations between $p_z$ and $p_y$ orbitals of Se atoms are dominating (Fig. 4(b)), the total MAE of MnSeI ML is contributed mainly by the orbital hybridizations of Se atoms. Based on the atomically orbital resolved MAE as shown in Fig. S6 in supplementary material, we obtain that chalcogen atoms Se in MnSeCl ML and halogen atoms $Y$ in MnS$Y$ ($Y$ = Br and I) MLs mainly determine their MAEs. It is worth noting that Lado *et al.* disclosed that the MAE of CrI$_3$ ML is mainly from the strong SOC of halogen atoms iodoine,[51] which is similar to MnS$Y$ ($Y$ = Br, I) MLs.

From the experimental perspectives, the synthetization of Janus Mn$XY$ should be feasible because the mixing of $X$ and $Y$ atoms in a disordered arrangement at its both surfaces can be prevented by controlling the growth conditions, as demonstrated in the successfully grown Janus MoSSe in a previous experiment.[22] On the other hand, Mn$XY$ is distinctly different from Janus Cr$XY$[25]. In the octahedral crystal field, Mn$^{3+}$ is a Jahn-Teller-active magnetic ion but Cr$^{3+}$ is not. Consequently, Mn$XY$ has strong JT distortions and its crystal structure is distinguished from that of Cr$XY$. Besides, Mn$XY$ is a FM half-metal with a high $T_C$ above room temperature but Cr$XY$ is a FM semiconductor with a $T_C$ no higher than 176 K or insulating magnet with spin spiral orders depending on $X$ and $Y$. As a result of the different electronic and magnetic properties of Mn$XY$ and Cr$XY$, the former is good sources of spin-flow injection[8] and has potential technological applications such as single-spin electron sources and high-efficiency magnetic sensors[52], whereas the latter is usually used to maximize the magnetic proximity effect through constructing heterostructures.

In summary, by means of first-principles calculations and MC simulations, we investigate structural, electronic and magnetic properties of 2D Janus magnets Mn$XY$ ($X$ = S and Se; $Y$ = Cl, Br and I) MLs. Through calculating their phonon spectra and performing AIMD simulations, we show that MnSBr, MnSI, MnSeCl and MnSeI MLs with JT distortions are thermodynamically stable. These four MLs are high-temperature ferromagnets with strong inplane MAEs and their $T_C$ can be as high as 418 K. Our calculated band structures show that they are intrinsic half-metals with spin band gaps larger than 2.4 eV. When SOC is considered, the 100% spin polarization at the Fermi level is not changed for MnSBr whereas the spin polarization at the Fermi level of other three MLs will be lower than 100%. We unveil that the strong MAEs of MnS$Y$ ($Y$ = Br and I) MLs mainly originates from halogen atoms while those of MnSe$Y$ ($Y$ = Cl and I)

MLs mainly comes from chalcogen atoms. These exciting findings in 2D Janus Mn$XY$ MLs present a type of above room temperature FM half-metals and provide opportunities for further spintronics applications.

See supplementary material for DFT computational and MC simulations details, structural information, phonon spectra of structures with space groups $P3m1$ and $Cm$, AIMD simulations results, magnetic moments, Heisenberg exchange parameters $J_{i\text{-}j}$ ($i$ = 1, 2, 3, 4 and 5; $j$ = 1, 2 and 3), the component values of DMI vector, the values of $|\boldsymbol{D}_{i\text{-}j,//}|$ ($i$ = 1; $j$ = 1 and 2), MAE, $T_\mathrm{C}$, the spin band gap, band structure and PDOS without (with) SOC, orbital-resolved MAE for Mn$XY$ MLs.

This work was supported by the National Key R&D Program of China (Grant No. 2022YFA1403301) and the National Natural Sciences Foundation of China (Grants No. 12104518, 92165204, 62104073 and U2130101), Natural Science Foundation of Guangdong Province (Grant No. 2022A1515012643, 2022A1515010035, 2023B1515120013). The DFT calculations reported were performed on resources provided by the Guangdong Provincial Key Laboratory of Magnetoelectric Physics and Devices (No. 2022B1212010008) and Tianhe-II.

AUTHOR DECLARATIONS

**Conflict of interest**

The authors have no conflicts to disclose.

**Author Contributions**

**Xiang-Fan Huang**: Investigation (equal); Methodology (equal); Writing – original (equal). **Kang-Jie Li**: Methodology (equal); **Zequan Wang**: Methodology (equal); **Shi-Bo Zhao**: Writing – review & editing (equal); **Bing Shen:** Writing – review & editing (equal); **Zu-Xin Chen**: Writing – review & editing (equal); **Yusheng Hou**: Conceptualization (equal); Funding acquisition (equal); Investigation (equal); Project administration (equal); Resources (equal); Supervision (equal); Writing – review &

editing (equal).

## DATA AVAILABILITY

The data that support the findings of this study are available from the corresponding authors upon reasonable request.